\documentclass{iopart}

\newcommand{\om}{\omega}

\newcommand{\be}{\begin{equation}}
\newcommand{\ee}{\end{equation}}
\newcommand{\bea}{\begin{eqnarray}}
\newcommand{\eea}{\end{eqnarray}}

\begin{document}

\comment[]{Comment on: ``Does an atom interferometer test the gravitational redshift at the Compton frequency?"}

\author{M A Hohensee,$^1$ S Chu,$^{1,2}$ A Peters$^3$ and H M\"uller$^1$}
\address{$^1$ Department of Physics, University of California, Berkeley, CA 94720, USA}
\address{$^2$ U.S. Department of Energy, 1000 Independence Ave., SW Washington, DC 20585, USA}
\address{$^3$ Humboldt-Universit\"at zu Berlin, Newtonstr. 15, 12489 Berlin, Germany}
\ead{hohensee@berkeley.edu}
\begin{abstract}
We show that Wolf {\em et al.}'s 2011 analysis in {\it Class. Quant. Grav.} {\bf 28,} 145017 does not support their conclusions, in particular that there is ``no redshift effect" in atom interferometers except in inconsistent dual Lagrangian formalisms. Wolf {\em et al.} misapply both Schiff's conjecture and the results of their own analysis when they conclude that atom interferometers are tests of the weak equivalence principle which only become redshift tests if Schiff's conjecture is invalid. Atom interferometers are direct redshift tests in any formalism.
\end{abstract}



Wolf, Blanchet, Bord\'e, Reynaud, Salomon, and Cohen-Tannoudji \cite{Wolf} give a very interesting discussion on 
comparing tests of the weak equivalence principle (WEP) and the gravitational redshift. They argue that atom interferometers (AIs) cannot be interpreted as redshift tests in any consistent theoretical formalism; the $7\times 10^{-9}$ precision of the AI test \cite{redshift} should not be viewed as a redshift test that is 10,000 or 280 times better than previous \cite{Vessot} or proposed \cite{ACES} experiments, but as a test of WEP.  Here, we show that while the analysis in the body of their paper is largely correct, it does not support these statements, made in their abstract and conclusion -- in particular, that AIs exhibit ``no redshift effect" and are not tests of the gravitational redshift unless interpreted in the context of an inconsistent dual Lagrangian formalism.\footnote{Moreover, \cite{Wolf} appears to assume that the interpretation was based on such a framework only in the first place. Instead, \cite{redshift} uses the atoms' acceleration of free fall $g'$ as an independent variable. It specifies explicitly that $g'$ may be 'derived from a principle of least action' using a single Lagrangian. Then, $g'=(1+\beta) g$, which amounts to the single-Lagrangian model of \cite{Wolf}. This has apparently been overlooked in \cite{Wolf}, see in particular p. 7 and the footnotes on pp. 4 and 12. By allowing $g'$ to take any other value, the analysis in \cite{redshift} also includes dual-Lagrangian formalisms.} Instead, their analysis reaffirms the fact that AIs are redshift tests in all frameworks.

Wolf {\em et al.} \cite{Wolf} first argue that the AI is not a clock comparison because the atom's Compton frequency $\om_C = mc^2/\hbar$ or equivalently its mass $m$ cancels out in the 
AI phase and is not measurable in the experiment.  We note that this is a generic feature of any measurement of the frequency ratio of two identical oscillators, as are common in redshift tests~\cite{Moriond}. The Pound-Rebka experiment \cite{PoundRebka,PoundSnider} is a good example: the redshift of a M\"ossbauer transition is deduced from the velocity $v=\Delta U/c$ of a M\"ossbauer absorber that cancels it by the first-order Doppler effect ($\Delta U$ is the gravitational potential difference). The frequency of the 14.4\,keV transition cancels out in this velocity and is not measurable in the experiment. Despite this, we are confident that the authors of \cite{Wolf} would agree that the Pound-Rekba experiment does measure the gravitational redshift. The absence of an explicit signal that is proportional to $\om_C$ is not evidence that AIs are not redshift tests.

Wolf {\em et al.} \cite{Wolf} then argue that the AI is not a redshift test because the action principle (or energy conservation) requires the free evolution phase accumulated between the matter waves to vanish for a closed interferometer. 
Specifically, in the laboratory frame,\footnote{The redshift must be referred to the same frame as $\Delta U$.} the AI phase shift is written (see (1.1-1.3) in \cite{Wolf}) as the sum
\bea
\om_C \oint d\tau+\phi_{\rm laser}=\phi_r+\phi_t+\phi_{\rm laser}, \nonumber \\  \phi_r=\om_C \int \frac{\Delta U}{c^2} dt, \quad \phi_t=-\om_C\int \frac{v_1^2-v_2^2}{2c^2 } dt
\eea
of three terms, respectively due to redshift, time dilation, and laser-atom interaction, where $\tau$ is the proper time and $v_{1,2}$ the velocities of the atom in the interferometer arms. We first note that the free evolution phase $\om_C\oint d\tau=\phi_r+\phi_t$ does not vanish in all closed interferometers, a fact used in AI determinations of the fine structure constant \cite{Biraben}. Wolf {\em et al.} \cite{Wolf} are correct that the effects of the redshift and time dilation sum to zero for the AI used in \cite{redshift}, $\phi_r+\phi_t=0$. This, however, is only possible if $\phi_r$ is nonzero (unless one wants to argue that there is no time dilation, $\phi_t=0$).  The redshift phase is recovered without further assumptions on the atom's interaction with gravity:
Based only on the fact that an atom's wavefunction accumulates phase in its rest frame at the Compton frequency (originally established by de Broglie \cite{deBroglie}) and energy-momentum conservation in non-gravitational interactions, a principle common to every theoretical framework \cite{Wolf,NatureReply,Nordtvedt,Ni,Giulini,redshiftPRL}, the time dilation phase $\phi_t$ cancels the laser phase $\phi_{\rm laser}$, whether or not that framework violates other laws of nature.\footnote{Assuming that the atoms follow parabolic trajectories with coordinate acceleration $dv/dt=-g'$ due to gravity, where $g'$ is arbitrary and need not be known, $\phi_t=-2\om_C \int_0^T (g'v_r t/ c^2)dt$, with $v_r$ equal to the atoms' recoil velocity after absorbing a photon of wavenumber $k$ and $T$ the pulse separation time.  If furthermore $v_r=\hbar k/m,$ this term must always be $-kg'T^2=-\phi_{\rm laser}$, where $\phi_{\rm laser}=\om_C v_r[x_1(T)+x_2(T)-x(0)-x(2T)]/c^2=kg'T^2$ is given by the atom's positions $x_{1,2}(t)$. The AI phase is thus given by the redshift term: $kgT^2.$} This applies in particular to both the single and dual-Lagrangian frameworks given in \cite{Wolf}. The AI used in \cite{redshift} is thus unambiguously a redshift test in all frameworks; the claim \cite{Wolf} of ``no redshift effect" in the AI confuses the free evolution phase with the redshift phase, which does not vanish.

Furthermore, Wolf {\em et al.} \cite{Wolf} make a subtle 
error about concepts that had been developed during the Schiff-Dicke controversy \cite{Schiff,Dicke}, in particular what is now known as Schiff's conjecture \cite{Willbook}. This conjecture proposes that any complete, self-consistent ({\em e.g.}, energy conserving, quantum-mechanics compatible) theory of gravity that satisfies WEP will also satisfy the full Einstein equivalence principle (EEP), which guarantees a conventional gravitational redshift. Indeed, constraints on anomalous redshifts can be translated to constraints on violations of WEP, provided that energy is conserved in gravitational interactions \cite{Schiff,Dicke,Willbook,Nordtvedt}.\footnote{Ni has shown that it is possible to violate EEP without producing signals accessible to torsion balances \cite{Ni}, but such violations always generate signals in experiments with freely falling bodies.}  
This is explicit in (3.20) where \cite{Wolf} shows that redshift and WEP tests are (sometimes differently) sensitive to the same EEP-violating parameter. This is confused for evidence that AIs are WEP tests. This confusion is understandable, as more traditional redshift tests focus on the small part of the particle's mass energy represented by the clock transition, and thus have different sensitivity to WEP and redshift anomalies. In contrast, the AI constrains redshift anomalies associated with the atom's full mass energy and thus provides equally strong constraints on WEP and redshift violation in theories which satisfy Schiff's conjecture. Far from proving that the AI is a WEP test and not a redshift test, part 1 of \cite{Wolf} merely reaffirms this relationship.  This agrees with \cite{redshiftPRL}, which shows that all tests of redshift or WEP are sensitive to linear combinations of the same parameters of the standard model extension \cite{Datatables}.  To argue, as \cite{Wolf} does on p. 5 and footnote 8, that AIs are not redshift tests simply because they bound the same parameters as WEP tests is inconsistent with Schiff's conjecture.

While we can use a redshift test to constrain WEP in theories that satisfy Schiff's conjecture, we cannot (by definition) do so in theories that violate it.  It is thus significant that Wolf {\em et al.} find the AI to be a redshift test in the context of their multiple-Lagrangian formalism that, aside from being inconsistent with quantum mechanics and energy conservation,\footnote{Such inconsistencies would pose problems for any test of the redshift or WEP, since all experiments, particularly atomic clocks and AIs, depend upon quantum mechanics or energy conservation at some level.  This of course is the main message conveyed by Schiff's conjecture: We know of no fully consistent theory of gravity in which the redshift is not inextricably linked to WEP.} also violates Schiff's conjecture.  The AI can only constrain redshift violations in such a theory if it does so directly, since WEP tests cannot do so indirectly.  Since the experiment is the same, regardless of which framework we analyze it in, this means that the AI must also be a direct test of the redshift in theories that do satisfy Schiff's conjecture.\footnote{An unambiguous test of WEP for matter-waves can be obtained by using fluorescence detection to trace the wave-packet's center of mass as a function of time (rather than splitting and interfering the wave packet). In theories that satisfy Schiff's conjecture, the resulting WEP violation constraint implies an equivalent limit on gravitational redshift anomalies experienced by matter-wave Compton clocks.  Such a matter-wave WEP test would not constrain redshift anomalies produced by theories that violate Schiff's conjecture, such as \cite{Wolf}'s multiple Lagrangian formalism.} This analysis \cite{Wolf} shows that the AI is a redshift test in both the single and dual Lagrangian formalisms. Wolf {\em et al.}, however, argue that the AI is fundamentally a WEP test which can be converted into a redshift test by relaxing Schiff's conjectured link between the redshift and WEP. This is logically impossible, and is a major problem with their paper's argument.

Wolf {\em et al.}'s remark  (p. 10) on a hypothetical redshift test with clocks based on matter/antimatter annihilation is also at odds with their conclusions. Suppose, {\em e.g.,} that we annihilate particle-antiparticle pairs to produce photon pairs at a frequency of $\om_C=mc^2/\hbar$. A cavity could in principle be used to store these photons. A continuous supply of pairs would create a ``Compton-laser" producing a macroscopic electric field. No-one will doubt that the oscillations of this field may be used as a clock. One may send one such clock each along two paths in a gravitational field and subsequently compare their phases. The measured difference is given by the integrals $\int \om_C d\tau$ taken along the two paths, {\it i.e.,} by the redshift and time dilation $\phi_r+\phi_t$, exactly as given above (1). To perform a pure redshift measurement, we may track the paths of the clocks to calculate $\phi_t$ for subtraction \cite{Vessot,ACES}.  This experiment is indeed not feasible with current technology (though existing atomic clocks could be used to perform a similar one with reduced sensitivity). The AI \cite{redshift}, however, is a feasible experiment that measures exactly the same phase: Instead of converting the particles to photons whose phases are compared, it directly compares the particles' phases by interference, but the result must be the same as long as the conversion into photons is coherent. The free evolution phase is given by $\phi_r+\phi_t$, and the laser-atom interaction effectively tracks the paths to subtract $\phi_t$. This, and the entire analogy, holds in general relativity as well as any alternative theory, provided only that nongravitational interactions conserve energy-momentum. Since nothing in \cite{Wolf} contradicts this, and since \cite{Wolf} agrees that matter/antimatter clocks could be used for redshift tests, we should conclude that the AI is a redshift test.

The other arguments in part 1.2 and elsewhere \cite{Wolf} are easily addressed: (i) The quan\-tum uncertainty principle precludes us from knowing which of two available tra\-jecto\-ries in the AI the atom takes (p. 5). It does not, however, prevent us from knowing the details of the available paths. In particular, the gravitational redshift is given by the vertical separation of the trajectories, which is known from the photon momentum and the atom's time of flight, provided only that energy-momentum is conserved \cite{JMO}. Similarly, there is no need to continuously track the atoms so as to guard against the effects of WEP violation, since for all theoretical frameworks yet considered, the atoms' vertical separation is independent of the direct effects of WEP violation. (ii) Possible non-quadratic terms in the Lagrangian (due, {\it e.g.,} to gravity gradients) prevent us from exactly calculating the phase by integrating over the classical path (pp. 5, 12). In the present situation of near-homogenous gra\-vi\-ta\-tio\-nal fields (as in any {\it local} test of general relativity), however, they are small enough to be treated perturbatively and of no con\-sequence \cite{JMO,Dimopoulos}. (iii) The AI can be used to measure the acceleration of free fall $g$ and its signal would vanish in free fall (p. 11). This, however, is true of any gravitational experiment in a homogenous gravitational potential, ({\it e.g.}, \cite{PoundRebka,PoundSnider}) and has no bearing on whether it is a redshift test.

Finally, \cite{Wolf} notes that in specialized scenarios, AIs may be much less sensitive than traditional clock comparisons (p. 18). We agree. The specific model \cite{Wolf}, however, is likely ruled out by relative redshift tests: anomalous redshifts between a Cs hyperfine clock and a H-maser \cite{Ashby}, an optical Sr clock \cite{Blatt}, and a Hg clock \cite{Fortier} have been pro\-bed with 1.4\,ppm, 3.5\,ppm, and 3.5\,ppm accuracy, respectively. Without fortuitous cancelations of redshift anomalies in these three experiments, the anomaly proposed in \cite{Wolf} is ruled out at the accuracy of proposed spaceborne atomic clock tests \cite{ACES}.

In summary, the claims made in the abstract and conclusion of \cite{Wolf} are not supported by their analysis, and indeed run counter to everything that is currently understood about gravitation, Schiff's conjecture, the gravitational redshift, and WEP.  Although the AI redshift test of \cite{redshift} may be used to validate WEP in the context of theories that satisfy Schiff's conjecture, it is fundamentally a redshift test, and is such in any theoretical framework.

\newpage

\end{document}